\documentclass[sn-basic]{sn-jnl}

\usepackage{graphicx}%
\usepackage{multirow}%
\usepackage{amsmath,amssymb,amsfonts}%
\usepackage{amsthm}%
\usepackage{mathrsfs}%
\usepackage[title]{appendix}%
\usepackage{xcolor}%
\usepackage{textcomp}%
\usepackage{manyfoot}%
\usepackage{booktabs}%
\usepackage{algorithm}%
\usepackage{algorithmicx}%
\usepackage{algpseudocode}%
\usepackage{listings}%
\usepackage{longtable}
\usepackage{adjustbox}
\usepackage{tabularx}

\begin{document}

\title[ ]{ A Study on the Evolutionary Period Changes of Short-period Type II Cepheids}


\author*[1,4,5]{\fnm{Yacob} \sur{Alemiye Mamo}}\email{malemiye@gmail.com}

\author[2]{\fnm{Berdinkov} \sur{N. Leonid}}
\email{lberdnikov@yandex.ru}
\equalcont{These authors contributed equally to this work.}

\author[3]{\fnm{Pastukhova} \sur{N. Elena}}
\email{pastukhova@sai.msu.ru }
\equalcont{These authors contributed equally to this work.}

\affil*[1]{\orgdiv{Astronomy and Astrophysics Department, Entoto Observatory and Research Center (EORC)}, \orgname{ Space Science and Geospatial Institute (SSGI) }, \orgaddress{\street{Menelik II Avenu}, \city{Addis Ababa}, \postcode{541}, \state{Addis Abab city Admin}, \country{Ethiopia}}}

\affil[2]{\orgdiv{ Sternberg Astronomical Institute}, \orgname{ Moscow State University }, \orgaddress{\street{Universitetskiy Prospekt, 13}, \city{Moscow}, \postcode{119234}, \state{ Moscow State}, \country{Russia}}}

\affil[3]{\orgdiv{ Institute of Astronomy}, \orgname{ Russian Academy of Sciences }, \orgaddress{\street{Pyatnitskaya ul. 48 }, \city{Moscow}, \postcode{119017}, \state{ Moscow State}, \country{Russia}}}

\affil[4]{\orgdiv{Physics Department}, \orgname{Addis Ababa University}, \orgaddress{\street{King George VI St }, \city{ Addis Abab}, \postcode{1176}, \state{Addis Abab city Admin}, \country{Ethiopia}}}

\affil[5]{\orgdiv{Leiden Observatory}, \orgname{Leiden University}, \orgaddress{\street{NL-2300 RA Leiden }, \city{ Leiden }, \postcode{9513}, \state{South Holland}, \country{The Netherlands}}}


\abstract{We investigate the period changes of 13 short-period Type II Cepheids using the O-C method over a century-long data baseline. The O-C diagrams for these stars exhibit a parabolic trend, indicating both increasing and decreasing period changes over time. These observed period changes are consistent with recent theoretical models based on horizontal branch evolutionary models for short-period BL Her stars. The pulsation stability test proposed by Lombard and Koen confirms that the period changes are significant, indicating evolutionary shifts. We identify seven BL Her stars with decreasing periods, expanding the existing sample size of short-period Type II Cepheids. This contributes to a deeper understanding of stellar evolution and the processes governing low-mass stars.}




\maketitle

\section{Introduction}\label{sec1}

Type II Cepheids (T2Cs) are low-mass stars (approximately 0.5$-$0.6 $ M_{\odot}$ ) exhibiting radial pulsations. In the Hertzsprung-Russell (HR) diagram, they are situated between classical Cepheids and RR Lyrae stars. Notably, in the HR diagram for the Large Magellanic Cloud (LMC), RR Lyrae stars and T2Cs form a single sequence. T2Cs have luminosities that are lower than those of classical Cepheids but higher than RR Lyrae stars \citep{Wallerstein2002}.These stars are crucial for distance measurements due to their unique Period-Luminosity relationship, providing insights into stellar evolution and galactic structure \citep{Catelan2015} .

T2Cs occupy the post-horizontal branch evolutionary phase, characterized by hydrogen and helium shell burning in low-mass stars. Based on their pulsation periods, these stars are categorized into three types: BL Herculis (BLH) stars, with periods between 1 and 4 days; W Virginis (WVs) stars, with periods between 4 and 20 days; and RV Tauri (RVs) stars, with periods over 20 days \citep{Soszynski2018}. Each group represents a distinct stage in post-horizontal branch evolution, enabling comparisons between theoretical predictions and observational data at each stage.

BLH stars,with short perods, cross the instability strip (IS) and show increasing periods as they evolve from the horizontal branch toward the Asymptotic Giant Branch (AGB). WVs stars, with intermediate periods, evolve at higher luminosities on the AGB and experience helium shell flashes. These flashes cause temporary movement from the AGB into the IS, leading to either increasing or decreasing periods. The long-period RVs stars, in the post-AGB phase, undergo thermal pulses that shift them blueward into the IS, resulting in consistently decreasing periods\citep{Gingold1976,Bono1997b,Wallerstein2002,Bhardwaj2022}. 


In this work, we focus on short-period Type II Cepheids (BL Herculis, or BLH stars) and emphasize period changes and the evolutionary status of these stars. Theoretical studies suggest that as BLH stars evolve from the hot to the cool edge of the instability strip (IS), their periods generally increase \citep{Gingold1976,Bono2016,Sandage1994,Wallerstein2002}. While the primary  evolutionary trend for BLH stars is redward evolution with increasing periods, studies by \cite{Gonzalez1994}and \cite{Yacob2022} have reported that some BLH stars exhibit decreasing periods. The current theoretical framework by \cite{Bono2020} predicts that BLH stars can experience both positive and negative period changes during post-horizontal branch evolution. A recent study by \cite{Yacob2022} (hereafter referred to as Paper I) also found both increasing and decreasing periods over a baseline of more than a century.

In this work, we extend our previous study (Paper I) by increasing the sample size to 13 and analyzing the period changes and evolutionary status of BLH stars with periods between 1 and 4 days. The 13 BLH stars included in this study are IU Cen, V553 Cen, BB Gem, V1149 Her, V447 Mon, V913 Mon, V2511 Sgr, FY Vir, ASAS094844-3921.4, ASAS092701+1412.1, ASAS140143-4650.8, ASAS142357-0100.1, and ASAS144155-1938.1.
The paper is organized as follows: Section \ref{Sec:sec2} outlines the data sources and methods, Section \ref{Sec:sec3} discusses the results, and Section \ref{Sec:sec4} concludes with a summary.

%

\section{Data, Methods and Analysis}\label{Sec:sec2}

This study utilized data from various ground-based and space-based surveys, covering $B$, $V$, and $g$ band observations, as well as relevant literature that are suitable for this study. Photographic data, PG,  were obtained from the Digital Access to a Sky Century at Harvard (DASCH)~ \citep{Grindlay2012} and the Sternberg Astronomical Institute (SAI). CCD data were sourced from the All-Sky Automated Survey (ASAS-3)~\citep{Pojmanski2002}, the All-Sky Automated Survey for Supernovae (ASAS-SN) ~\citep{Kochanek2017}, INTEGRAL-OMC ~ \citep{AlfonsoGarzon2012}, Catalina Sky Survey (CSS) ~\citep{Drake2013}, PAN-STARRS ~ \citep{Chambers2016}, and HIPPARCOS ~\citep{Perryman1997} for $V$ and $g$ band measurements. Additional photoelectric (PHE) data with $B$ and $V$ filters, as well as photovisual (PGV) data from various literature sources, were also utilized. Table~\ref{tab:tabel1} provides the sources of data, detailing the observation period, the number and type of observations, and the references for each target star in the study.

Data from multiple sources were compiled and organized chronologically using Heliocentric Julian Date (HJD). These data were grouped into seasonal batches, typically spanning at least three months per year, with intervals adjusted based on data availability. Light curves were constructed for each star, and $O-C$  (Observed minus Calculated) diagrams were created to detect period changes by comparing observed times of maximum light with calculated times. T2Cs often exhibit asymmetric light curves. For our analysis, we selected times of maximum light (Tmax), as BL Her stars typically have sharper, well-defined maxima compared to their minima or midpoints, enabling more precise long-term measurements.

The analysis employed the \cite{Hertzsprung1919} method, using an algorithm by \cite{Berdnikov1992} to determine  $O-C$ residuals and correct for time shifts across different observation filters. The Hertzsprung's method utilizes the entire light curve to derive $O-C$  points for a variable star by fitting the phase of a light template (with its maximum shifted to phase 0.0) to the observations. Additionally, the \cite{Lombard1993} was applied to validate observed pulsation period changes.

To quantify period changes, the study used quadratic elements based on Sterken's approach \citep{Sterken2005}. These elements calculated the times of maximum light ($HJD_{\textrm{max}}$) using a quadratic equation (Eqn. 1), which includes terms for the period at a reference epoch $ M_{\textrm{o}}$, the period itself ($P$), a parabolic term $Q$, and the cycle number ($E$). The quadratic coefficient ($Q$) was essential for measuring the rate of period change ($dP/dt$)in units of seconds per year ($s/yr$) or days per million years ($d/Myr$)  as in Eqn. 3 and Eqn. 4 respectively, using sideral years, relative to the mid-epoch period ($P_{\textrm{mid}}$) and cycle number ($E$). The coefficient $Q$ is related to $O-C$ as shown in Eqn. 2.

\begin{align}
\label{eq:1}
HJD_{\textrm{max}} &= M_{\textrm{o}} + PE + QE^{2} , \\
\label{eq:2}
Q &= \frac{1}{2}\frac{dp}{dt}P_{\textrm{mid}}E^2, \\
\label{eq:3}
\frac{dP}{dt} &= 365.25\times24\times60\times60\times(\frac{2Q}{P_{\textrm{mid}}})  , \\
\label{eq:4}
\frac{dP}{dt} &= 730.5\times10^{6}\times(\frac{Q}{P_{\textrm{mid}}}).
\end{align}

\begin{longtable}{@{}lllll@{}}	
	\caption{Summary of the Data Sources}
	\label{tab:tabel1} \\
 	\hline
 	\textbf{Star Name} & \textbf{Julian Date(JD)\&} & \textbf{No. of} & \textbf{Type of} &  
 	    \textbf{References}  \\
			&     \textbf{Year Intervals}          &  \textbf{Observ.}  & 	\textbf{Observation} & \\
 	\noalign{\smallskip} \hline \noalign{\medskip}                                                                                                                                                                                                                                                              
 	\endfirsthead
	 \multicolumn{5}{c}%
		{{\bfseries Table \thetable\ Summary of data Source ..continued }} \\
 	\hline
 	\textbf{Star Name} & \textbf{Julian Date(JD)\&} & \textbf{No. of} & \textbf{Type of} &  
 	    \textbf{References}  \\
			&     \textbf{Year Intervals}           &  \textbf{Observ.}  & 		\textbf{Observation} & \\
	 \noalign{\smallskip} \hline \noalign{\medskip}                                                                                                                                                                                                                                                              
	\endhead
	\hline
	\endfoot
	\hline
\endlastfoot

 			IU Cen    & 2411191 - 2458590 & 1117         & PG                   & DASCH                                               \\
			&     \textbf{\textcolor{blue}{1889 - 2019[130yrs]} }  & 553  & CCD ($V$) & INTEGRAL- OMC                                                                                                                                     \\
			&                      & 176                     & PHE ($B$,$V$)      & \begin{tabular}[c]{@{}l@{}}\cite{Walraven1958}\\\cite{Berdnikov2023}\\\cite{Mitchell1964} \end{tabular}\\

			&                      & 1032                   & CCD ($g$,$V$)           & ASAS-SN                                                                                                                                          \\
			&                      & 706                    & CCD ($V$)                & ASAS3               \\
			
			V553 Cen             & 2411168  -2458589    & 2794    & PG                 & DASCH                                                                                                                                              \\
			&  \textbf{\textcolor{blue}{1889 - 2019[130yrs]} }                    & 15                     & CCD ($V$)                & INTEGRAL-OMC                                                                                                                                      \\
			&                      & 152                    & PHE ($B$,$V$)    & \begin{tabular}[c]{@{}l@{}} \cite{wisse1970}\\\citet{Gray1991}\\ \cite{Harris1980};\cite{cotrell1979}\\ \citet{Dean1981};\citet{Diethelm1986}\\ \cite{Eggen1985}\\ \cite{Dean1977}\\ \citet{lloyd1972}\end{tabular} \\

			&                      & 1040                   & CCD ($g$,$V$)           & ASAS-SN                                                                                                                                          \\
			&                      & 541                    & CCD ($V$)                  & ASAS3                                                                                                                                           \\
			BB Gem             & 2411378 -2458588    & 1903                   & PG                           & DASCH                                                                                                        \\
			&   \textbf{\textcolor{blue}{1890 - 2019[129yrs]}}    & 234      & PG                                                         & \begin{tabular}[c]{@{}l@{}}\cite{Kinman1965};\cite{Zonn1935}\\ \cite{Borzdyko1964}\\ \cite{Satyvaldiev1970} \\ \cite{Teplitskaya1951}\end{tabular}\\

            &                      & 162                       & PG                 & SAI, this work
\\
			&                      & 10                       & CCD ($g$)                 & PAN-STARSS                                                                                                                                       \\
			&                      & 118                     & CCD ($V$)                  & INTEGRAL-OMC                                                                                                                                     \\
			&                      & 189                     & PHE ($B $,$V$)    & \begin{tabular}[c]{@{}l@{}} \cite{Bersier1994} \\ \cite{Oosterhoff1960}\\ \cite{Mitchell1964}\\ \cite{Berdnikov2023}\\  \cite{Henden1980}\\ \cite{Szabados1977} \end{tabular}\\

			&                      & 876                   & CCD ($g$,$V$)           & ASAS-SN                                                                                                                                           \\
           &                      & 364                    & CCD ($V$)                  & ASAS3                                                                                                                                           \\
			V1149 Her            & 2411954  - 2458590   & 456   & PG         & DASCH                                   \\
			&  \textbf{\textcolor{blue}{1891 - 2022[131 yrs]} } & 213 & PG           & SAI this Work                                                                                                                                             \\	
			&                      & 10                      & CCD($g$)      & PAN-STARSS                                                                                                                                     \\
			&                      & 15                      & CCD (V)      & INTEGRAL-OMC                                                                                                                                     \\

			&                      & 1647                    & CCD ($g$,$V$)           & ASAS-SN                                                                                                                                          \\
			&                      & 646                     & CCD ($V$)                  & ASAS3                                                                                                                                            \\
			V447 Mon              & 2413482-2458589   & 110    & PG            & DASCH                                                       \\
			&   \textbf{\textcolor{blue}{1895-2022[127 yrs]} }  & 109 & PG   & SAI this work 
\\
            &                      & 12              & CCD($g$)        & PAN-STARSS                                                                                                                                     \\
            &                      & 14              & CCD ($V$)       & INTEGRAL-OMC                                                                                                                                     \\
			&                      & 61                      & PHE ($B$,$V$)  & \begin{tabular}[c]{@{}l@{}} \cite{Harris1980};\cite{Henden1996} \\ \cite{Pont1997}\\ \cite{Berdnikov2023} \end{tabular}\\

           &                      & 900           & CCD ( $g$,$V$ )         & ASAS-SN                                                                                                                                           \\
			&                      & 157               & CCD($V$)             & ASAS3                                                                                                                                              \\		
			
			V913 Mon           & 2411749-2459770      & 1102       & PG         &	DASCH                                                                                                                                            \\
			&  \textbf{\textcolor{blue}{1891-2022[131 yrs]}}  & 63 & PG   & SAI this work                                                                                                                     \\
			&                      & 5944                  & CCD ($g$,$V$)    & ASAS-SN                                                                                                                                          \\
            &                      & 1194                  & CCD($V$)          & ASAS3                                                                                                                                              \\			
			V2511 Sgr  &  2411237  -2458590  & 340         & PG                 & DASCH                                                                                                                                           \\
			& \textbf{\textcolor{blue}{1889 -2019[130 yrs]}} & 213 & PG & SAI this work 
\\
					&                & 10               & CCD ($g$)        & PAN-STARSS                                                                                                                                       \\
			&                      & 15              & CCD ($V$)          & INTEGRAL-OMC                                                                                                                                     \\
			&                      & 152            & PHE ($B$,$V$)   & \begin{tabular}[c]{@{}l@{}} \cite{Chambers2016}\\ \cite{Berdnikov2023} \end{tabular} \\
      
			&                      & 646             & CCD($g$,$V$)     & ASAS-SN                                                                                                                                           \\
	
			FY Vir           & 2437047-2458305     & 22    & PG           & DASCH                                                                                                                                             \\
	      & \textbf{\textcolor{blue}{1905-2019[114yrs]}}  & 104    & PG & \begin{tabular}[c]{@{}l@{}} \cite{Goranskij1979}\\ \cite{Berdnikov2023}\end{tabular}\\

			&                      & 12           & CCD($g$)           & PAN-STARSS                                                                                                                                    
 \\
			&                      & 3141          & CCD ($V$)           & INTEGRAL-OMC                                                                                                                                   
 \\
 			&                      & 41            & PHE ($B$,$V$)    & \begin{tabular}[c]{@{}l@{}} \cite{Chambers2016}\\\cite{Henden1996} \end{tabular}\\

			&                      & 426             & CCD ($V$)           & CSS                                                                                                                                              \\
			&                      & 1126           & CCD ($g$,$V$)        & ASAS-SN                                                                                                                                         \\
         ASAS094844-3921.4    & 2412601-458590   & 659   & PG           & DASCH                                                                                                                                             \\
	      & \textbf{\textcolor{blue}{1905-2019[114yrs]}}  & 2974   & PHE ($B$,$V$)    & \cite{Berdnikov2023}
\\
			&                      & 176            & CCD ($V$)           & CSS                                                                                                                                              \\
			&                      & 1126           & CCD ($g'$,$V$)        & ASAS-SN                                                                                                                                         \\
	  
          ASAS092701+1412.1    & 2412146-2458587  & 797  & PG              & DASCH                                                                                                                                             \\
	      & \textbf{\textcolor{blue}{1892-2019[127yrs]}}  & 14   & CCD($g$) & PAN-STARSS                                                                                                                                    
\\
	        &                      & 304            & PHE ($B$,$V$)    & \begin{tabular}[c]{@{}l@{}} \cite{Chambers2016}\\ \cite{Berdnikov2023}\end{tabular}\\                                                                               	
             &                      & 708           & CCD ($V$)           & CSS
\\                                                                                                                         
			&                      & 1751          & CCD ($g'$,$V$)        & ASAS-SN                                                                                                                                         \\
			 	  
	  ASAS140143-4650.8        & 2411194-2458590 & 1100  & PG           & DASCH                                                                                                                                             \\
	      & \textbf{\textcolor{blue}{1889-2022[133 yrs]}}   & 268   & PHE ($B$,$V$)  & \cite{Berdnikov2023}
\\
			&                      & 106            & CCD ($V$)           & CSS                                                                                                                                              \\
			&                      & 1086          & CCD ($g$,$V$)        & ASAS-SN                                                                                                                                        
 \\
            &                      & 88           & CCD ($g$,$V$)        & ASAS3                                                                                                                     
 \\
	    ASAS142357-0100.1       & 2412973-2458591 & 834  & PG           & DASCH                                                                                                                                             \\
	      & \textbf{\textcolor{blue}{1894-2019[125 yrs]]}} & 219   & PHE ($B$,$V$)  & \cite{Chambers2016}
\\
			&                      & 15           & CCD($g'$)           & PAN-STARSS                                                                                                                                    
 \\
			&                      & 412            & CCD ($V$)           & CSS                                                                                                                                              \\
			&                      & 2339          & CCD ($g$,$V$)        & ASAS-SN                                                                                                                                         \\
		
			ASAS144155-1938.1       & 2411155 - 2458590 & 1861 & PG           & DASCH                                                                                                                                             \\
	      & \textbf{\textcolor{blue}{1889-2019[133 yrs]}} & 287   & PHE ($B$,$V$)  & \cite{Berdnikov2023}
\\
			&                      & 11           & CCD($g$)           & PAN-STARSS                                                                                                                                    
 \\
			&                      & 329            & CCD ($V$)           & CSS                                                                                                                                              \\
			&                      & 1531          & CCD ($g$,$V$)        & ASAS-SN                                                                                                                                         \\							

\end{longtable}	


\begin{figure}
   	\centering
  	\includegraphics[width=\textwidth]{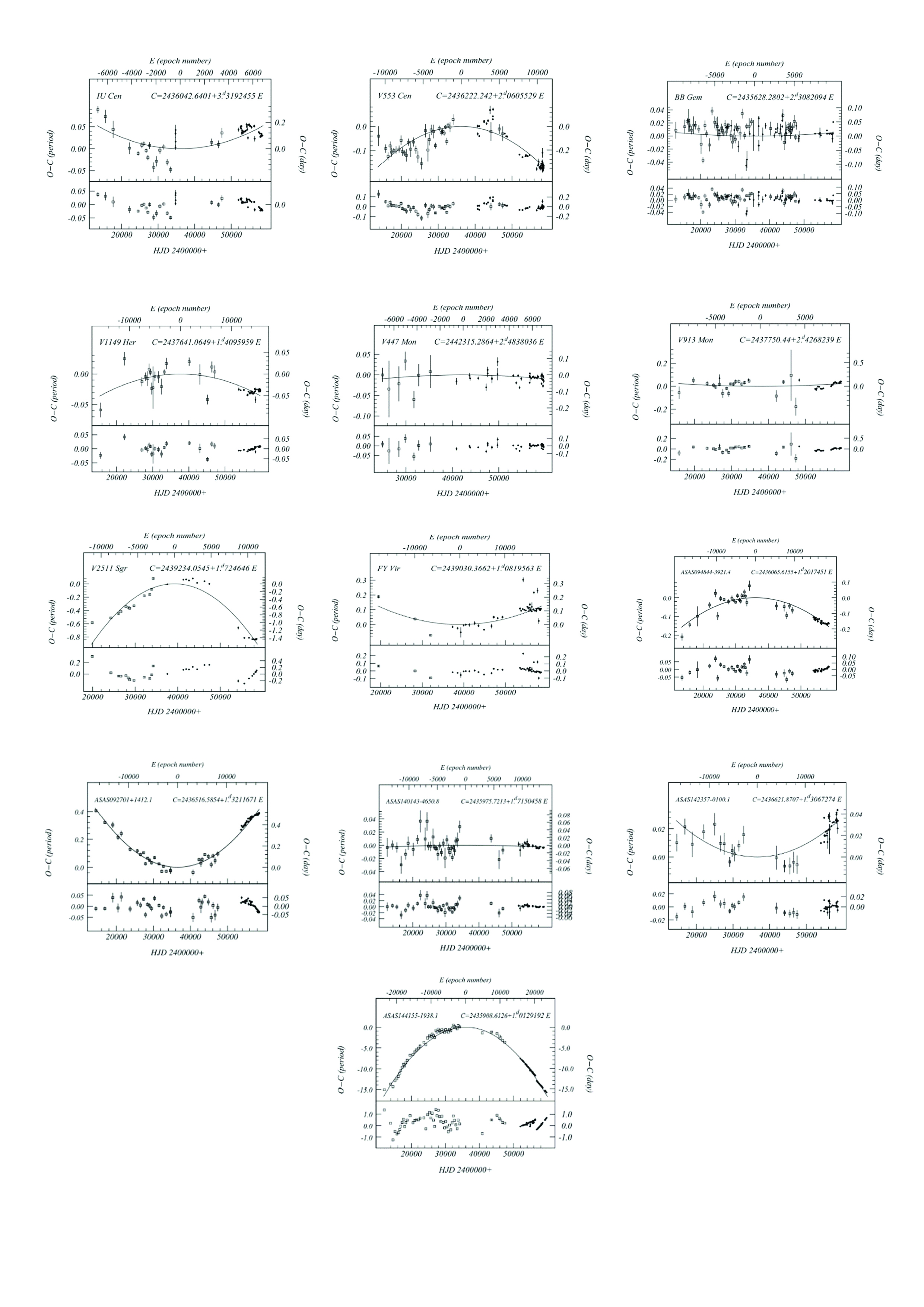}
\caption{The $O-C$ diagrams for 13 stars relative to the linear (top panel) and quadratic (bottom panel) elements (Table \ref{tab:table3} ). The open squares represent the Harvard PG observations, while dots are used for all other measurements; vertical bars indicating the limits of errors in the residuals. }\label{fig:OC}  
   \end{figure}
   
\begin{figure}
   	\centering
  	\includegraphics[width=\textwidth]{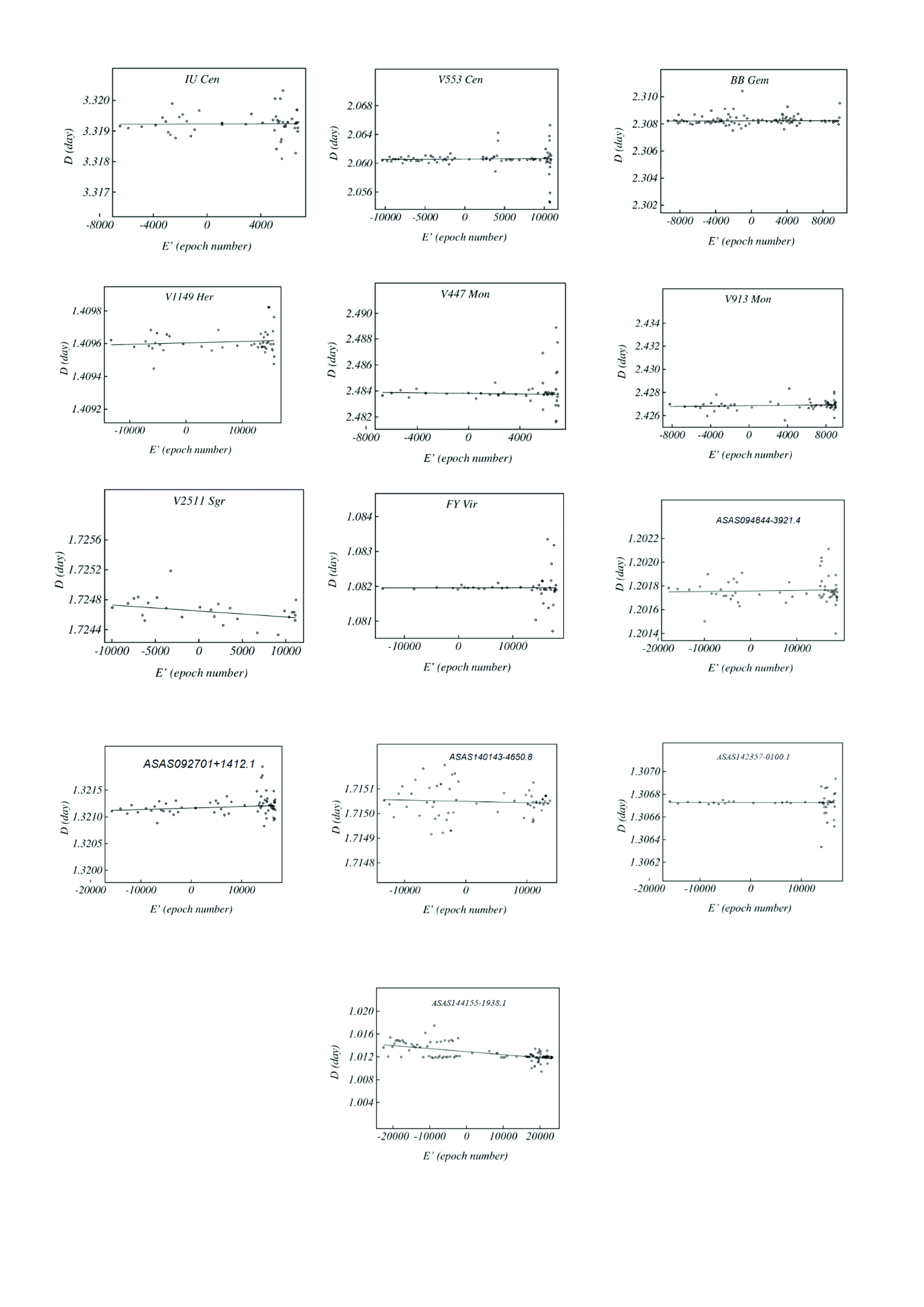}
\caption{The period change in epoch interval ($D$ ) in days versus the epoch number ($E$), showing the period change corresponding to the $O-C$ residuals. }\label{fig:KOEN}  
   \end{figure}

\section{Results and Discussion}\label{Sec:sec3}

The $O-C$ diagrams for the 13 BL Her stars reveal parabolic trends, indicating continuous period changes. Among these, six stars exhibit increasing periods, consistent with theoretical redward evolution. Conversely, seven stars display decreasing periods, suggesting potential blueward evolution. These findings challenge traditional models \citep[e.g.,][]{Gingold1976} and align with recent theoretical frameworks predicting mixed period behaviors \citep[e.g.,] [] {Bono2020, Yacob2022}.The period change test confirms these trends as significant, providing robust evidence for their evolutionary nature.

The O-C diagrams for each star (Figure \ref{fig:OC}) display deviations of the observed times of maximum light from the calculated times based on a constant period. The diagrams exhibit both concave-upward and concave-downward parabolas, indicating increasing and decreasing period changes,respectively. A parabolic shape in the O-C diagram signifies a continuous period change. Open squares represent Harvard photographic measurements, while dots denote other observations, with vertical bars indicating the error limits in the O-C residuals. The parabolic elements for each star are listed in Table \ref{tab:table2}. The quadratic coefficient Q from Table \ref{tab:table2} is used to calculate the observed rate of period change (dP/dt), which is listed in column 5 of Table \ref{tab:table3}, along with its associated error.

Among the four pass bands used ($V$, $B$ and $PG$, and $g$), $V$-band serves as the primary and corrections were made to adjust the $O-C$ diagrams for the $B$ and $g$ bands (Paper I and \citet{Arp1955}). These corrections for the $B$ and $g$ bands are provided in Table \ref{tab:table3} and were applied in constructing Figure 1 and determining the elements in Table \ref{tab:table3}.

The $O-C$ diagram results revealed a parabolic evolutionary trend. Among the 13 BL Her stars, six stars (IU Cen, BB Gem, FY Vir, V913 Mon, ASAS142357-0100.1, and ASAS092701+1412.1) with concave-upward parabolas show increasing periods, while seven stars (V553 Cen, V1149 Her, V447 Mon , V2511 Sgr, ASAS094844-3921.4, ASAS140143-4650.8, and ASAS144155-1938.1) with concave-downward parabolas exhibit decreasing period changes. 

  	\begin{table}
		
		\caption{The quadratic elements }
		\label{tab:table2}
			\begin{tabular}{|c|l|} 
				\hline
				\textbf{ Stars} & \textbf{$HJD_{max} = M_{\textrm{o}} + PE  +  QE^{2}$}, where $M_{\textrm{o}}$ \textbf{\textcolor{blue}{adopted epoch }}, \\
				&  $P$ \textbf{\textcolor{blue}{is the period at  adopted epoch }}and  $Q$ is parabolic term \\[0.5ex]
				\hline
				IU Cen    & $36042.6401  + 3.3192455 E  +  0.34152\times 10^{-10}E^{2}$ \\
				V553 Cen  & $36222.2419  + 2.0605529 E  - 0.29213\times 10^{-8} E^{2}$ \\
				BB Gem    & $35628.2802  + 2.3082094 E  - 0.12573\times 10^{-9} E^{2}$\\
				V1149 Her  & $237641.0649 + 1.4095959 E  - 0.204\times 10^{-9} E^{2}$\\
				V447 Mon  & $42315.2864  + 2.4838036 E  - 0.4738\times 10^{-9} E^{2}$\\
				V913 Mon  & $37750.4399  + 2.4268239 E  + 0.68740\times 10^{-9} E^{2}$\\
				V2511 Sgr & $39234.0545  + 1.724646  E  + 0.12243\times 10^{-7} E^{2}$\\
				FY Vir    & $39030.3662  + 1.0819563 E  + 0.40964\times 10^{-9} E^{2}$\\ [1ex]    
				ASAS094844-3921.4  & $36065.6255 + 1.2017451 E - 0.5336\times 10^{-9} E^{2}$\\
				ASAS092701+1412.1  & $36516.5854 + 1.321167 E - 0.19813\times 10^{-8} E^{2}$\\
			ASAS140143-4650.8  & $35975.7213 + 1.7150458 E -0.25628\times 10^{-10} E^{2}$\\
				ASAS142357-0100.1 & $36621.8707 + 1.3067274 E  +  0.118\times 10^{-9} E^{2}$\\
			ASAS144155-1938.1 & $35908.6126 + 1.0129192 E - 0.30135\times 10^{-7} E^{2}$\\[1ex]    
				\hline     
			\end{tabular}
		
	\end{table}

\begin{table}
\caption{Corrections for maxima in the $B$ and $ g'$ band, New Element and rate of period change($dp/dt$)}\label{tab:table3}
\begin{tabular}{@{}lllll@{}}
\toprule%
Stars             & Corrections for & Corrections for  & New elements            & Rates of period \\
                 & maxima in $B$   &  maxima in $g $   &                         &  change $dp/dt$   \\
                  & in days(d) & in days(d) &     &($s/yr$)  \\
\midrule
IU Cen            & 0.00886                     & -0.02091                     & 36042.6401 + 3.3192455E & 0.07060 $\pm$  0.01431    \\
V553 Cen          & 0.04079                     & 0.11249                      & 36222.2419 + 2.0605529E & -0.08948 $\pm$  0.00643   \\
BB Gem            & 0.00256                     & 0.01487                      & 35628.2802 + 2.3082094E & 0.00344 $\pm$ 0.00299    \\
V1149HER          & -                           & -0.00809                     & 37641.0649 + 1.4095959E & -0.00913 $\pm$  0.00195 \\
V447 Mon          & -0.01675                    & -0.03807                     & 42315.2864 + 2.4838036E & -0.01204 $\pm$ 0.00884   \\
V913 MON          & -                           & -0.00460                     & 37750.4399 + 2.4268239E & 0.01788 $\pm$  0.01743    \\
V2511 Sgr         & 0.02019                     & -0.00532                     & 39234.0545 + 1.724646E  & -0.44804 $\pm$  0.03046   \\
FY Vir            & 0.01247                     & 0.00552                      & 39030.3662 + 1.0819563E & 0.02390 $\pm$  0.00589    \\
ASAS094844-3921.4 & -0.01088                    & -0.00494                     & 36065.6255 + 1.2017451E & -0.02798 $\pm$  0.00204     \\
ASAS092701+1412.1 & 0.00997                     & -0.00854                     & 36516.5854 + 1.321167E  & 0.09449 $\pm$  0.00243    \\
ASAS140143-4650.8 & 0.00246                     & -0.00417                     & 35975.7213 + 1.7150458E & -0.00094 $\pm$  0.00163   \\
ASAS142357-0100.1 & -0.00248                    & -0.01324                     & 36621.8707 + 1.3067274E & 0.00570$\pm$ 0.00086      \\
ASAS144155-1938.1 & -0.05379                    & 0.02288                      & 35908.6126 + 1.0129192E & -1.87576$\pm$ 0.02164 \\
\noalign{\smallskip}
\hline  
\end{tabular}
\end{table}

\subsection{Stars with Increasing Periods}\label{subsec1}
Among the 13 BLH stars studied, six show (IU Cen, BB Gem, FY Vir, V913 Mon, ASAS142357-0100.1, and ASAS092701+1412.1) concave-upward parabolas in their $O-C$ diagrams, indicative of increasing periods. This result is consistent with the expectations from theoretical models that predict redward evolution as BLH stars evolve from the hot to the cool edge of the instability strip \citep{Gingold1976,Sandage1994}. The observed increasing periods align with previous findings by \cite{Yacob2022}, who reported similar trends in their smaller sample of BLH stars. 
While the overall trend for these stars is consistent with previous studies, there is a notable variation in the rate of period change. For example, IU Cen and BB Gem show moderate increases in their periods, while FY Vir and V913 Mon exhibit a slightly steeper rate of increase. These differences in the rate of period change may reflect variations in the stars' mass, metallicity, or other stellar parameters, as suggested by previous studies \citep{Bono2016}.

\begin{table}
\caption{Summary of Rate of period change for the 13 short T2CS (BLHs)}\label{tab:table4}
\begin{tabular}{@{}llllll@{}}
\toprule
Stars             & Period     & $dp/dt$    & Error ($s/yr$) & Remark of period & Year of Span \\
\midrule
IU Cen            & 3.31924554 & 0.07060  & 0.01431          & \textbf{\textcolor{red}{Increasing}}     & 130          \\
BB Gem            & 2.30820941 & 0.00344  & 0.00299          & \textbf{\textcolor{red}{Increasing}}       & 129          \\
FY VIR            & 1.0819563  & 0.02390  & 0.00589          & \textbf{\textcolor{red}{Increasing}}       & 114          \\
V913 Mon          & 2.42682393 & 0.01788  & 0.01743          & \textbf{\textcolor{red}{Increasing}}       & 131          \\
ASAS142357-0100.1 & 1.30672765 & 0.00570  & 0.00086          & \textbf{\textcolor{red}{Increasing}}       & 125          \\
ASAS092701+1412.1 & 1.32116709 & 0.09449  & 0.00243          & \textbf{\textcolor{red}{Increasing}}       & 127          \\
V553 Cen          & 2.06055294 & -0.08948 & 0.00643          & \textbf{\textcolor{blue}{Decreasing}}     & 130          \\
V1149 Her          & 1.40959589 & -0.00913 & 0.00195          & \textbf{\textcolor{blue}{Decreasing}}       & 131          \\
V447 Mon          & 2.48380356 & -0.01204 & 0.00884          & \textbf{\textcolor{blue}{Decreasing}}       & 127          \\
V2511 Sgr         & 1.724646   & -0.44804 & 0.03046          & \textbf{\textcolor{blue}{Decreasing}}       & 130          \\
ASAS094844-3921.4 & 1.20174485 & -0.02798 & 0.00204          & \textbf{\textcolor{blue}{Decreasing}}      & 126          \\
ASAS140143-4650.8 & 1.71504581 & -0.00094 & 0.00163          & \textbf{\textcolor{blue}{Decreasing}}      & 133          \\
ASAS144155-1938.1 & 1.01191373 & -1.87576 & 0.02164          & \textbf{\textcolor{blue}{Decreasing}}       & 133         \\   
\botrule  
\end{tabular}  
\end{table}

\subsection{Stars with Decreasing Periods}\label{subsubsec2}
 A key finding of this study is the identification of seven stars (V553 Cen, V1149 Her, V447 Mon , V2511 Sgr, ASAS094844-3921.4, ASAS140143-4650.8, and ASAS144155-1938.1) exhibiting concave-downward parabolas, indicating decreasing periods. This result challenges the traditional model of BLH stars evolving redward with increasing periods. Decreasing periods had been observed in a few BLH stars in previous studies, such as  \cite{Gonzalez1994} and \cite{Yacob2022}, but the present study significantly expands the sample size, providing stronger statistical evidence for this phenomenon.
 
 The decreasing periods observed in these stars suggest that some BLH stars may experience blueward evolution, moving across the instability strip in a way that is not fully explained by existing theoretical models. This observation aligns with the theoretical framework proposed by \cite{Bono2020}, which suggests that short-period BLH stars may experience both positive and negative period changes during their post-HB evolution. Our findings add empirical evidence to support this theoretical model.
 
 This study advances our understanding of BL Her stars by demonstrating both increasing and decreasing period changes, offering insights into their complex evolutionary pathways. These findings support recent models of post-horizontal branch evolution, highlighting the importance of long-term monitoring in stellar astrophysics. Future research should aim to integrate more extensive datasets and refine theoretical models to better understand these evolutionary processes.

\subsection{Validation of Period Changes}\label{subsubsec3}

To validate period changes in BLH stars,we employed the \cite{Lombard1993} test, which distinguishes between systematic (evolutionary) and random (stochastic) variations. This method is particularly well-suited for analyzing long-term period changes driven by stellar evolution. In contrast, the Eddington-Plakidis test detects random cycle-to-cycle fluctuations, which are more relevant for stars like RR Lyrae and classical Cepheids. BLH stars ($1-4$ days) primarily exhibit gradual evolutionary period changes, making the Lombard and Koen test more suitable for our study of long-term trends (Paper I).

To assess the significance of the period changes, we used the error estimates associated with the quadratic coefficient ($Q$). A period change is considered significant if the slope ($dP/dt$) exceeds three times its standard error (3\(\sigma\) criterion).

The Lombard and Koen method confirmed that the observed period changes are significant and not due to random fluctuations or measurement errors. The test graphs (Figure \ref{fig:KOEN}) show that the period changes in most stars exhibit clear and consistent trends over time, further supporting the evolutionary nature of these period changes. This validation is crucial as it strengthens the reliability of our findings and establishes the observed period changes as real, rather than artifacts of observational uncertainties or random noise.

While abrupt period changes are observed in T2Cs, our analysis focuses on long-term, gradual period variations over a century-long baseline. The uncertainties in the determined period change rates ($dP/dt$), derived from the errors in the quadratic coefficient ($Q$), are listed in Table \ref{tab:table4}. These uncertainties typically range from 0.001 to 0.03 s/year and account for both observational errors and potential small-scale abrupt variations in the O?C diagrams. This ensures that the reported trends reflect genuine evolutionary changes rather than short-term fluctuations.
 
The $O-C$ diagrams reveal a mixture of increasing and decreasing period trends, consistent with theoretical predictions. The presence of both types of behavior further supports the hypothesis that short-period BLH stars can undergo both positive and negative period changes during their evolution, as suggested by recent theoretical models \citep{Bono2020}. These results also provide observational evidence for the presence of Gravo-Nuclear Loops (GNLs), proposed by \citet{Bono1997a,Bono1997b}, offering insights into the helium-burning mixing process in low-mass stars. This process is crucial for our understanding of stellar evolution. Additionally, our findings increase the existing sample size of short-period T2Cs with period-decreasing behavior.
%
\section{Summary and Conclusions }\label{Sec:sec4}

This study advances our understanding of BL Her stars by demonstrating both increasing and decreasing period changes, offering insights into their complex evolutionary pathways (Tabel \ref{tab:table4}). These findings support recent models of post-horizontal branch evolution, highlighting the importance of long-term monitoring in stellar astrophysics. Future research should aim to integrate more extensive datasets and refine theoretical models to better understand these evolutionary processes.
\begin{itemize}
  	\item  The calculated rates of period change (dP/dt) for the 13 BL Her stars vary, showing both increasing and decreasing trends.
  	\item  The O-C diagrams display clear parabolic trends for most stars, confirming continuous period changes.
   	\item  The period change test graphs validate these changes, reinforcing the reliability of our results.
\end{itemize}  	
   	
This study advances the understanding of period changes in short-period Type II Cepheids (BL Her stars) by expanding the sample size and utilizing a century-long data baseline. Notably, seven stars showed decreasing periods, contrary to the traditionally expected redward evolution with increasing periods, providing empirical evidence for Gravo-Nuclear Loops (GNLs) as predicted by recent evolutionary models. These findings are relevant to understanding the helium-burning mixing process, which is crucial for our understanding of stellar evolution. These results offer deeper insights into the post-horizontal branch evolution of low-mass stars and highlight the importance of long-term monitoring in stellar astrophysics.

Overall, this research extends previous work on BL Her stars by increasing the sample size and offering a detailed analysis of period changes. The findings align with theoretical predictions and contribute new insights into the evolutionary status of short-period Type II Cepheids.

\backmatter

\bmhead{Supplementary information}

The times of maximum light, derived from the reduced seasonal light curves for all stars, are presented in Table A1 at appendices as Supplementary Information, which is crucial for the work of this paper. 

%
%

\bmhead{Acknowledgements}

We thank the anonymous referee for valuable comments that improved the paper significantly. AMY acknowledges and thanks the Leiden Observatory, in Leiden, The Netherlands, for the continuous support provided by research administrators and staff during his research visit at Leiden Observatory. AMY also acknowledges the Lomonosov Moscow State University, where the study was conducted in collaboration with the Sternberg  Astronomical Institute (SAI), for providing the data crucial to this work. Finally, AMY is deeply indebted to the Space Science and Geospatial Institute (SSGI) in Ethiopia for for their unwavering support throughout this research study.
This work makes use of the observational data available from the catalogue of Digital Access to a Sky Century at Harvard (DASCH), All Sky Automated Surveys (ASAS-3), All Sky Automated Surveys of SuperNovae (ASAS-SN), INTEGRAL-OMC, Catalina Sky Survey (CSS), PAN-STARSS, HIPPARCOS and data from Sternberg, Astronomical Institute (SAI).
\bmhead{Data Availability}
The data underlying this article are available in the article as well as the data source cited in the article and references therein.
\bmhead{Author contributions}
AMY undergone the study, wrote manuscript, and reduced and analyzed the data. LNB reviewed and edit the manuscript beside provide all the algorithm developed for data reduction, ENP collected some part of the data which is crucial for the study.
\bmhead{Competing interests}
The authors declare no competing interests.
\bibliography{sn-bibliography}%
%
%

%
%
%
%

\begin{appendices}

\section{Times of maximum light}\label{secA1}
Table A1 presents the observed times of maximum light. Columns 1-3 provide the times of maximum light in Modified Julian Day (MJD) format, their corresponding calendar year, and the associated measurement errors. Column 4 specifies the type of observations used. Columns 5 and 6 display the epoch number (E) and the O-C residual. Columns 7 and 8 indicate the number of observations used for each measurement and their respective data sources.

	




\end{appendices}



\end{document}